\documentclass[onecolumn,showpacs,preprintnumbers,elsart]{revtex4}
\usepackage{makeidx}
\usepackage{amssymb}
\usepackage{amsmath}
\usepackage{mathrsfs}
\usepackage{graphicx}
\usepackage{dcolumn}
\usepackage{bm}
\usepackage[center]{subfigure}
\usepackage{color}

\begin{document}

\title{Dipolar matter-wave solitons in two-dimensional anisotropic discrete lattices}
\author{Huaiyu Chen$^{1}$, Yan Liu$^{1}$, Qiang Zhang$^{1}$, Yuhan Shi$^{1}$, Wei Pang$^{2}$, and Yongyao Li$^{1}$}
\email{yongyaoli@gmail.com}
\affiliation{
$^{1}$Department of Applied Physics, College of Electronic Engineering, South China Agricultural University,
Guangzhou 510642, China \\
$^{2}$ Department of Experiment Teaching, Guangdong University of
Technology, Guangzhou 510006, China.}

\begin{abstract}
We numerically demonstrate two-dimensional (2D) matter-wave solitons in the disk-shaped dipolar Bose-Einstein condensates (BECs) trapped in strongly anisotropic optical lattices (OLs) in a disk's plane. The considered OLs are square lattices which can be formed by interfering two pairs of plane waves with different intensities. The hopping rates of the condensates between two adjacent lattices in the orthogonal directions are different, which gives rise to a linearly anisotropic system. We find that when the polarized orientation of the dipoles is parallel to disk's plane with the same direction, the combined effects of the linearly anisotropy and the nonlocal nonlinear anisotropy strongly influence the formations, as well as the dynamics of the lattice solitons. Particularly, the isotropy-pattern solitons (IPSs) are found when these combined effects reach a balance. Motion, collision and rotation of the IPSs are also studied in detail by means of systematic simulations. We further find that these IPSs can move freely in the 2D anisotropic discrete system, hence giving rise to an anisotropic effective mass. Four types of collisions between the IPSs are identified. By rotating an external magnetic field up to a critical angular velocity, the IPSs can still remain localized and play as a breather. Finally, the influences from the combined effects between the linear and the nonlocal nonlinear anisotropy with consideration of the contact and/or local nonlinearity are discussed too.
\end{abstract}

\maketitle



\section{Introduction}
\begin{figure}[tbp]
\centering%
{\label{fig1a}
\includegraphics[scale=0.4]{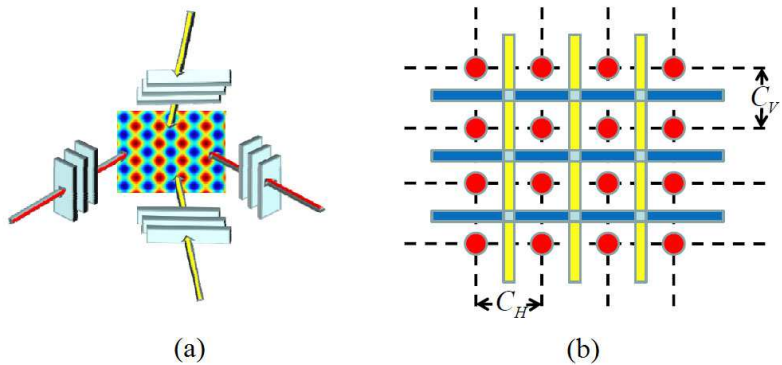}}%
\caption{(Color online) (a)An experimental sketch for realizing 2D optical lattices by interfering two pairs of plane waves with different intensities in each pair. (b) The obtained 2D anisotropic lattices. Due to the different intensities in each pair of the plane waves, the hopping rates between the adjacent lattices in the horizontal and vertical directions are different.}
\label{Fig1}
\end{figure}

Bose-Einstein condensates (BECs) of atoms and/or molecules which carry large magnetic or electric moments have attracted much attention in scientific research. Experimentally, BECs have been created in the ultracold gas of chromium \cite{Griesmaier}, in dysprosium atoms, in erbium atoms \cite{Aikawa,MLu}, etc. The creation of electric-dipolar BECs in the molecular gas \cite{Englert,Salhi} might also be expected. A review of this rapidly progressing field was given in Ref. \cite{Lahaye}. Theoretically, remarkable findings related to BECs are yielded, such as the formations of various kinds of BEC patterns \cite{Saito,Nath}, the prediction of stable dipolar BECs trapped in optical-lattice (OL) potentials \cite{Maluckov,Klawunn}, the roton instability of BECs \cite{Wilson,Hufnagl}, and the demonstration of the possibility of the Einstein-de Haas effect \cite{Gawryluk}.

Dipole-dipole interactions (DDIs) are crucial for systematically studying the properties of the dipolar BECs. Ultracold dipolar quantum gas has been proven to be an ideal simulator for studying the nonlocally nonlinear effect, which is resulted from the DDIs between the polar atoms and/or molecules. Experimental studies demonstrated that the sign of DDIs can be changed by a rotating magnetic field \cite{Giovanazzi}. Further, when the dipole polarization is parallel to the two-dimensional(2D) plane with the same polarized direction, the nonlocal nonlinearity becomes anisotropic.  An interesting ramification of such studies is the prediction to the 2D solitons since stable 2D bright solitons could be found in the background of nonlocal nonlinearity \cite{Skupin,Krolikowski}. In the past decade, the solutions of stable 2D bright, three-dimensional(3D) dark, and vortex bright solitons were found with the consideration of isotropic DDIs \cite{Pedri,Nath2,Tikhonenkov}. Particularly, the repulsive DDIs induced by nonuniform dc fields in the gas of the polarizable particles might give rise to a completely different type of bright solitons \cite{Yongyao1}. Stable 2D anisotropic fundamental solitons were also proved to be able to exist in the dipolar BECs when dipoles are parallel to the 2D plane \cite{Tikhonenkov2,Raghunandan}. Very recently, stripe solitons and anisotropic vortex solitons have been shown to be able to exist stably in dipolar BECs with the consideration of the spin-orbit coupling \cite{YPZ,Xunda}. These findings allow a new avenue for studying the physics of higher dimensional solitons. Meantime, dipolar matter-wave solitons in 1D settings also attract much attention \cite{Cuevas,Yongyao2,Bland}.

Usually, the trapping of BECs \cite{Brazhnyi,Morsch,ZWXie,McKay,Hamner}, as well as the supporting of abundant matter-wave gap or discrete solitons \cite{Trombettoni,Alfimov,efremidis,Ostrovskaya,Louis,Eiermann,Ahufinger,Kevrekidis2005-2,Baizakov,Sakaguchi,Yongping2} could be realized in optical lattices (OLs). A natural extension is to study the effect of OL potential on the soliton formations in the dipolar BECs, in which case the discrete model with the DDIs between remote lattice sites should be considered\cite{Maluckov2}. It was shown that the DDIs can enhance the stability of discrete solitons. For example, stable unstaggered 1D \cite{Gligoric,Gligoric2}, 2D bright \cite{Gligoric3} discrete solitons and the discrete vortex solitons \cite{Gligoric4} were shown to be supported by the DDIs. Moreover, it was shown that the motion of the discrete solitons across the lattice can be controlled by the polarized orientation of the dipole moment. This is because the attractive DDIs compensate the Peierls-Nabarro (PN) potential barrier between the lattices \cite{Gligoric,Kivshar,Morandotti}, while the repulsive DDIs enhance the PN potential barrier \cite{Gligoric,Zhihuan}. In 2D space, this effect could also lead to anisotropic motion of the discrete solitons across the 2D lattices \cite{Yongyao3}. We note that the studies reporting the discrete soliton formations are mostly based on the isotropic OL potential. As a general case, OLs with different potential strength and/or periods, namely anisotropic OLs, are rarely considered \cite{Kevredkidis2005,Mayteevarunyoo,CGH}. Therefore, it's quite natural to consider how the DDIs will affect the formation, stability and tunability of matter-wave solitons in 2D or 3D anisotropic OLs. We expect that with the consideration of the anisotropic OLs, this setting will generate more freedoms for controlling the matter-wave solitons.

In this work, we numerically investigate the 2D solitons in the disk-shaped dipolar BECs trapped in a strong anisotropic OL potential in the disk's plane. The OL is a square lattice, which can be formed by interfering two pairs of plane waves with different intensity [see Fig. \ref{Fig1}(a)]. Therefore, the hopping rates of the condensates between the adjacent lattices along the mutually perpendicular directions are different [see Fig. \ref{Fig1}(b)], which can be considered as a linearly anisotropic OLs. The objective of this work is to study the combined actions of both the linear and nonlinear anisotropic effects upon the solitons. The pattern formation, motion, collision and rotation of the solitons under such combined effects are studied. This paper is structured as follows, the model of the system is described in Sec. II, the patterns of the solitons in this system are studied in Sec. III, and numerical results of the motion, collision and rotation of the solitons are presented in Sec. IV. To study the combined effects solely from the anisotropic linearity and nonlocal nonlinearity, we fix the strength of the contact nonlinearity by 0 in these two sections. To study the influence of the contact nonlinearity, we reconsidered the model by adding the contact nonlinearity in Sec. V. The paper concludes in Sec. VI.

\section{The models}
The underlying 3D Gross-Pitaevskii equation (GPE), which considers both the contact and/or local nonlinearity and the DDIs
corresponding to the OL potential, can be reduced to its normalized 2D equation,
\begin{gather}
i\frac{\partial \psi }{\partial t}=-\frac{1}{2}\left( \frac{\partial ^{2}}{%
\partial x^{2}}+\frac{\partial ^{2}}{\partial y^{2}}\right) \psi
+g\left\vert \psi \right\vert ^{2}\psi -U\left( x,y\right) \psi   \notag \\
+c\psi (\mathbf{r})\int \left\vert \psi \left( \mathbf{r}^{\prime }\right)
\right\vert ^{2}\left( 1-3\cos ^{2}\theta \right) \frac{d\mathbf{r}^{\prime }%
}{\left\vert \mathbf{r}-\mathbf{r}^{\prime }\right\vert ^{3}},  \label{GP}
\end{gather}%
where $\mathbf{r}\equiv \left\{ x,y\right\} $, $\theta $ is the angle
between vector $(\mathbf{r}-\mathbf{r}')$ and the dipole orientation fixed by a strong external magnetic field. $U\left( x,y\right)=A_{1}\cos(kx)+A_{2}\cos(ky)$
represents the OL potential, with $k$ the period of the potential. $A_{1,2}$ represent the intensities of two plane waves.
The derivation of Eq. (\ref{GP}) assumes the factorization of the 3D\ mean-field wave function, $\Psi $, under the action of the tight trapping
potential imposed in the transverse direction:%
\begin{equation}
\Psi \left( X,Y,Z,T\right) =\left( \sqrt{\pi }a_{\perp }^{3}\right)
^{-1/2}\exp \left( -\frac{i\hbar }{2ma_{\perp }^{2}}t-\frac{1}{2}%
z^{2}\right) \psi \left( x,y,t\right) ,  \label{Psi}
\end{equation}%
where the scaled spatial and temporal coordinates $X,Y,Z,T$ are related to physical ones, %
as follows: $\left\{ X,Y,Z\right\} \equiv a_{\perp }\left\{ x,y,z\right\}
,~T\equiv \left( ma_{\perp }^{2}/\hbar \right) t$, where $~a_{\perp }$ is the
transverse localized length, and $m$ is the mass of the particle. Following the factorized ansatz (\ref{Psi}), the total number of particles in
the condensates is given by%
\begin{equation}
P_{\mathrm{total}}=\int \int \left\vert \psi \left( x,y\right) \right\vert
^{2}dxdy.  \label{N}
\end{equation}%
The rescaling implied above gives rise to the expressions of dimensionless
coefficients of the contact and long-range interactions, given by
\begin{equation}
g=2\sqrt{2\pi }\frac{a_{s}}{a_{\perp }},~~c=\frac{4m\wp^{2}}{3\hbar
^{2}a_{\perp }}.  \label{g-G}
\end{equation}
where $a_{s}$ is the scattering length, and $\wp$ the dipole moment. Assuming the size of the strong OL potential, whose period is $2\pi/k$, is essentially larger than the transverse size $a_{\perp}$ of the confined dipoles, the BECs split into droplets
coupled by the weak linear interactions, due to the tunneling effect of atoms across the potential barriers. According to the experimental
estimation in Refs. \cite{Gligoric,Gligoric2,Gligoric3,Gligoric4}, for chromium atoms, we have $a_{\perp}\sim0.5$ $\mu$m, and the period of OL can be
readily set as 5 $\mu$m. Therefore, in the case of the dipolar BECs, the DDI may be considered as the interaction between the dipolar droplets
with collective magnetic moments, trapped in the local potential wells of the OL. While inside the strongly confined droplets,
the DDI can be reduced to a contact form. Under this circumstance, the continuous equation (\ref{GP}) can be replaced by its discrete counterpart.
To this end, the continuous wave function is decomposed over a set of modes strongly localized in the vicinity of each OL site (Wannier functions),%
\begin{equation}
\psi \left( x,y,t\right) =\sum_{m,n}\psi _{mn}(t)\Phi _{mn}(x,y),
\end{equation}%
where $m$ and $n$ denote the discrete coordinates of the lattice, as was done in the
course of the derivation of the one-dimensional(1D) \cite{Gligoric} and 2D \cite{Gligoric3,Yongyao3}
discrete models for the dipolar BECs trapped in the deep OL potentials. The rescaled discrete version of Eq. (\ref{GP}), i.e., a 2D discrete GPE with long-range
interactions, is then given by %
\begin{gather}
\partial _{t}\psi _{mn}=-\frac{C_{\mathrm{H}}}{2}\left( \psi _{m+1,n}+\psi _{m-1,n}-2\psi _{mn}\right)-\frac{C_{\mathrm{V}}}{2}\left( \psi _{m,n+1}+\psi _{m,n-1}-2\psi _{mn}\right)\notag \\
+\sigma |\psi _{mn}|^{2}\psi _{mn}\notag\\
+\psi _{mn}\sum_{\left\{ m^{\prime },n^{\prime }\right\} \neq \left\{
m,n\right\} }f_{\mathrm{DD}}(m-m^{\prime },n-n^{\prime })|\psi _{m^{\prime
}n^{\prime }}|^{2},  \label{discr}
\end{gather}%
where $C_{\mathrm{H,V}}$ are the dimensionless hopping rates on the horizontal ($x$) and vertical ($y$) directions [see Fig. \ref{Fig1}(a)], respectively. We note that the hopping rates $C_{\mathrm{H,V}}$ that can be controlled by the intensities of two interfering plane waves, are proportional to the value of overlapping integral of the Wannier functions localized at the adjacent sites of the lattice. $f_{\mathrm{DD}}(m-m^{\prime },n-n^{\prime })$ is the discrete DDI kernel, written as
\begin{equation}
f_{\mathrm{DD}}(m-m^{\prime },n-n^{\prime })={\frac{1-3\cos^{2}\theta}{[(m-m^{\prime })^{2}+(n-n^{\prime })^{2}]^{3/2}}%
},  \label{QQInonlocal}
\end{equation}%
where
\begin{eqnarray}
\theta\left(m,m',n,n'\right)=\arccos\left[{(m-m')^{2}\over(m-m')^{2}+(n-n')^{2}}\right]^{1\over2}. \label{theta}
\end{eqnarray}
Here, we have assumed that the polarized direction of the dipoles are in-plane and parallel to the $x$ axis.

The rescaling performed here is in the same way as it was done in Ref. \cite{Gligoric3,Yongyao3}, using coefficients that are expressed in the form of normalization and overlapping integrals built of the Wannier functions. Note that the coefficient in front of the DDI terms is scaled here to be $1$, and $\sigma $ is the relative strength of the contact interactions. Obviously, Eq. (\ref{discr}) conserves the norm of the discrete wave
function, written as
\begin{equation}
P=\sum_{m,n}|\psi _{mn}|^{2}.  \label{P}
\end{equation}

In the following sections, we will numerically study the formation, motion, collision and rotation of the solitons, as well as the influence of the hopping rates $C_{\mathrm{H,V}}$ and the total norm $P$ on them. To study the combined effects of the anisotropic nonlocal nonlinearity and linearity clearly, we will fix $\sigma\equiv0$ in Sec. III and IV. The system with $\sigma\neq0$ is discussed in Sec. V.

\section{Numerical results}
\subsection{Patterns of solitons}
\begin{figure}[tbp]
\centering%
{\label{fig2a}
\includegraphics[scale=0.3]{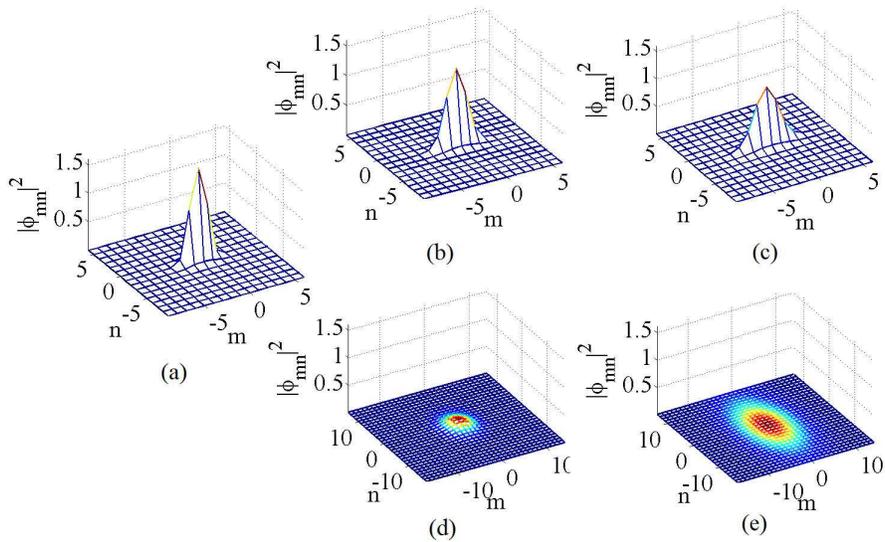}}%
\caption{(Color online) Typical examples of the stable fundamental solitons with $(C_{\mathrm{H}},C_{\mathrm{V}})=(1,1)$ (a), (3,1) (b), (5,1) (c), (1,3) (d) and (1,5) (e). In all panels, we have fixed $\sigma=0$, $P=4$, and assumed that all the dipoles are orientated to the $m$ (horizontal) direction. In simulations, the numerical domain is set as $128\times128$, and the regions of these panels are selected according to the size of the solitons. }
\label{Expsoliton}
\end{figure}

We first look for stationary fundamental (on-site and unstaggered) solutions of Eq. (\ref{discr}) with real chemical potential $\mu$. We assume that the solitary solutions have the following form,
\begin{equation}
\psi _{mn}(t)=\phi _{mn}e^{-i\mu t}.  \label{mu}
\end{equation}%
Stationary profiles $\phi_{mn}$ can be found in a finite domain by means of the well-known imaginary-time
propagation method \cite{Chiofalo,Jianke1,Jianke2}. Note that in this section, we have fixed the strength of the contact nonlinearity, $\sigma$, to be 0, and assumed that the polarization direction of all the dipoles are paralleled to the $m$ direction.

Typical examples of the fundamental (unstaggered bright) lattice solitons with different sets of $C_{\mathrm{H,V}}$ are displayed in Fig. \ref{Expsoliton}. The presented solitary solutions in Fig. \ref{Expsoliton} are stable (the stability of the solitons is tested by the real-time evolutions with sufficient evolved time). It is shown that the solitons in Fig. \ref{Expsoliton} are all on-site, thus called on-site solitons. We note that the profile of the soliton produced in isotropic lattice shown in Fig. \ref{Expsoliton}(a), is similar to its counterparts generated by the isotropic OLs with anisotropic DDIs \cite{Gligoric3}. The numerical results in Fig. \ref{Expsoliton} indicate that the selection of the values of $C_{\mathrm{H,V}}$ can not only change the size of the soliton, but also enhance, weaken, or even alter the anisotropic property of the soliton. Compared to Fig. \ref{Expsoliton}(a) in which case the lattice is isotropic, we find that increasing $C_{\mathrm{H}}$ can make the soliton become compressed in the $n$-direction, but prolate in the $m$-direction, then making it retreat to an 1D soliton. However, increasing $C_{\mathrm{V}}$ can make the soliton expanding in both directions, which leads it to a quasicontinued 2D soliton.

To further quantify the solitons, we introduce the effective area $S$ to describe the size of the soliton, and the parameter $\varepsilon$ to describe the degree of structural anisotropy of the solitons. The definition of these two parameters are given by
\begin{equation}
\mathrm{S}={\frac{\left( \sum_{m,n}|\phi _{mn}|^{2}\right) ^{2}}{%
\sum_{m,n}|\phi _{mn}|^{4}}};  \label{Aeff}
\end{equation}%
\begin{gather}
\varepsilon={\frac{D_{\mathrm{hor}}-D_{\mathrm{vert}%
}}{D_{\mathrm{hor}}+D_{\mathrm{vert}}}},~
\label{Anis} \\
D_{\mathrm{hor}}\equiv {\frac{\left( \sum_{m}|\phi _{m,0}|^{2}\right) ^{2}}{%
\sum_{m}|\phi _{m,0}|^{4}}},~D_{\mathrm{vert}}\equiv {\frac{\left(
\sum_{n}|\phi _{0,n}|^{2}\right) ^{2}}{\sum_{n}|\phi _{0,n}|^{4}}},\quad
\notag
\end{gather}%
Here, to clearly identify the $m$ and $n$ directions, we rename $m$ and $n$-directions as horizontal,``hor," and vertical, ``vert," directions, respectively.
\begin{figure}[tbp]
\centering%
{\label{fig3a}
\includegraphics[scale=0.33]{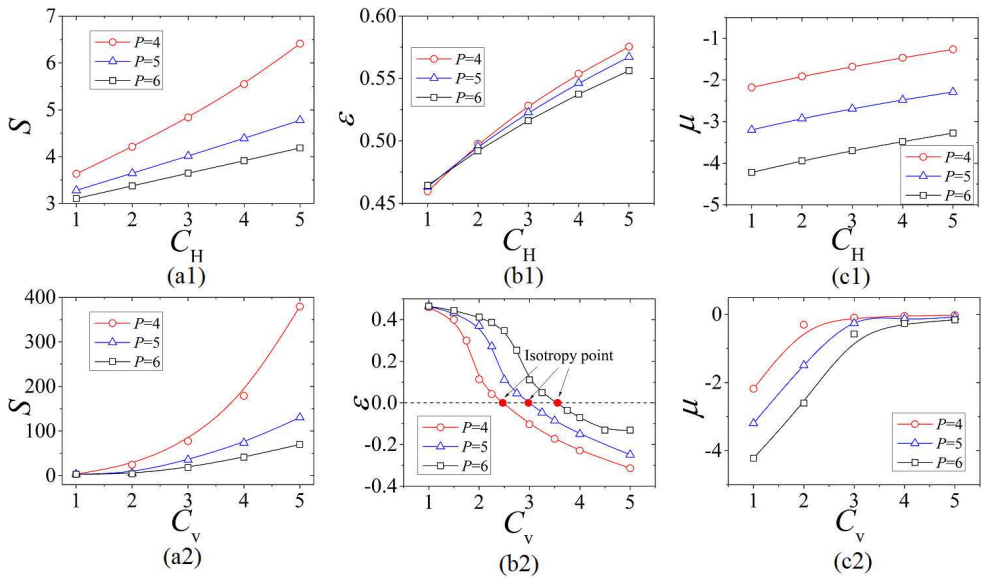}}%
\caption{(Color online) (a1,b1,c1) $S$, $\varepsilon$ and $\mu$ vs $C_{\mathrm{H}}$, respectively. Here, $C_{\mathrm{V}}\equiv1$. (a2,b2,c2) $S$, $\varepsilon$ and $\mu$ versus $C_{\mathrm{V}}$, respectively. Here, $C_{\mathrm{H}}\equiv1$. The dash line in panel (b2) shows $\varepsilon\equiv0$, in which case isotropic pattern of solitons can be found. All these figures are based on the conditions of $\sigma=0$, and that the dipoles in the lattices are orientated to the $m$ (horizontal) direction.}
\label{Char}
\end{figure}

In Eq. (\ref{Anis}), when $\varepsilon>0$, that means the horizontal length of the solitons is longer than the vertical length, i.e., alias $D_{\mathrm{hor}}>D_{\mathrm{ver}}$, so we name this type of solitons as horizontally-elongated solitons; when $\varepsilon<0$, that means the horizontal length of the solitons is shorter than the vertical length, i.e., alias $D_{\mathrm{hor}}<D_{\mathrm{ver}}$. In this case we name this type of solitons as vertically-elongated solitons; Particularly, if $\varepsilon=0$, i.e., $D_{\mathrm{hor}}=D_{\mathrm{ver}}$, the pattern of the solitons is isotropic. Therefore, the degree of structural anisotropy of solitons can be well described by $\varepsilon$.

Figure \ref{Char} clearly shows the dependence of the effective area $S$, the anisotropic degree $\varepsilon$, and the chemical potential $\mu$ on the hopping rates $C_{\mathrm{H,V}}$. Figures \ref{Char}(a1) and \ref{Char}(a2) show that increasing the hopping rate in both directions can also increase the effective area of the solitons, which can be naturally understood with the fact that increasing the hopping rate can also enhance the spreading effect of the solitons. We note that, the spreading effects of the solitons caused by the hopping rates in two directions are obviously different, as judged from Figs. \ref{Char}(a1) and \ref{Char}(a2). Specifically, increasing the hopping rate $C_{\mathrm{V}}$ (perpendicular to the dipole orientation) can lead to a stronger spreading effect than its orthogonal counterpart (i.e., $C_{\mathrm{H}}$). This phenomenon can be explained as following: in the horizonal direction, the spreading effect is partially suppressed by the attractive DDIs, while in the vertical direction, the spreading effect is enhanced by the repulsive DDIs. In Fig. \ref{Char}(b1), we find that the value of $\varepsilon$ grows monotonously as  $C_{\mathrm{H}}$ (here $C_{\mathrm{V}}\equiv1$). It indicates that the horizontal elongation of the soliton is enhanced. However, Fig. \ref{Char}(b2) shows that with the increasing of $C_{\mathrm{V}}$, the value of $\varepsilon$ first decreases, and then approaches to zero, after which it increases in its negative direction. It means that by increasing $C_{\mathrm{V}}$, solitons experience a transition from a horizontally elongated state (i.e. $\varepsilon>0$) to a vertically elongated one(i.e. $\varepsilon<0$). Based on the results shown in Figs. \ref{Char}(a1)-\ref{Char}(b2), we further find that solitons containing smaller norms are more sensitive to $C_{\mathrm{H,V}}$ than that of the larger norms. This phenomenon can be understood as the competition between the linear and the nonlinear effects: solitons containing larger norms have a stronger nonlinearity, which makes the solitons become less sensitive to $C_{\mathrm{H,V}}$. Figures \ref{Char}(c1) and \ref{Char}(c2) demonstrate the relationship between $\mu$ and $C_{\mathrm{H,V}}$. In these two panels, $\mu$ increases monotonously as $C_{\mathrm{H,V}}$, and we find that solitons with smaller norms have a larger value of $\mu$, satisfying the Vakhitov-Kolokolov (VK) criterion \cite{VK}, a necessarily stability condition of soliton in the attractive nonlinearity.

\begin{figure}[tbp]
\centering%
{\label{fig4a}
\includegraphics[scale=0.25]{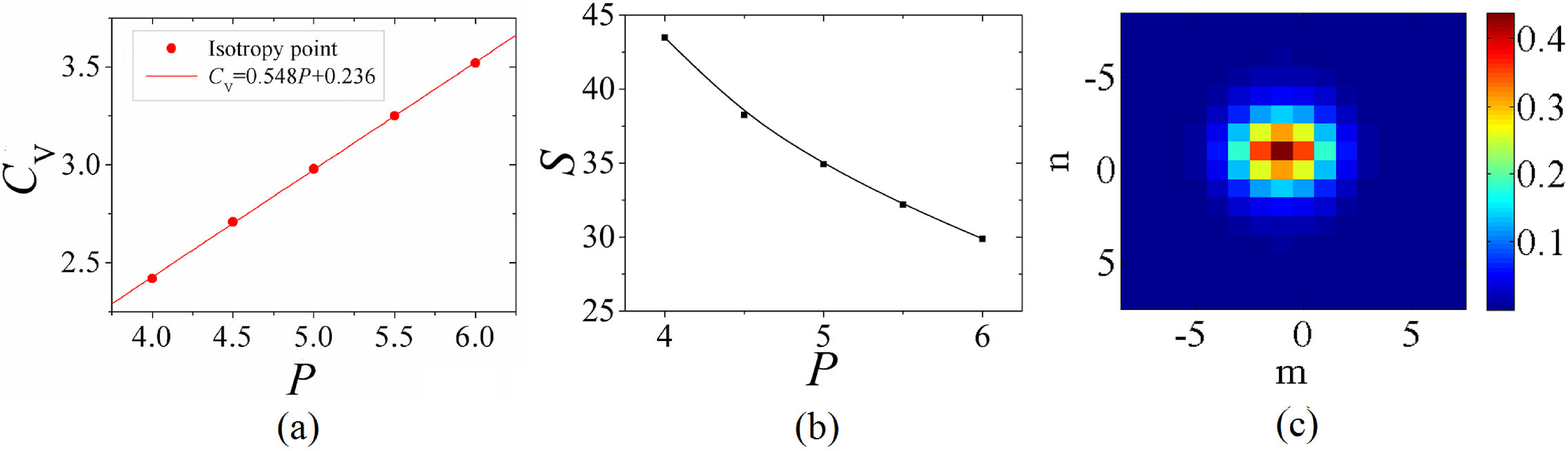}}%
\caption{(Color online) (a)The relationship between $C_{\mathrm{V}}$ and $P$, determined by the linearly fitted function $C_{\mathrm{V}}=0.548P+0.236$. Here, $C_{\mathrm{H}}\equiv1$. (b) The effective area of the IPSs vs $P$. (c) A typical example of an IPS, with the parameters setting as $(C_{\mathrm{H}},C_{\mathrm{V}},P)=(1,3.46, 6)$, and $\sigma=0$. We assume that the orientation of the dipoles are directed to the $m$ (horizontal) direction. }
\label{iso}
\end{figure}

\subsection{Isotropy-pattern soliton}
Particularly, in Fig. \ref{Char}(b2), when the value of $\varepsilon(C_{\mathrm{V}})$ approaches to $\varepsilon=0$, the patterns of solitons become isotropic, i.e. $D_{\mathrm{hor}}=D_{\mathrm{vert}}$. Numerical simulations further show that these isotropy-pattern solitons (IPSs) can be created on the condition of $C_{\mathrm{V}}=0.548P+0.236$ [see the red line in Figure \ref{iso}(a) for a fixed $C_{\mathrm{H}}\equiv1$]. Fig. \ref{iso}(a) further suggests that vertically-elongated solitons can be found within the regime above the red line; while the horizontally-elongated solitons exhibit inside the regime that is below the red line. We also study the effective area of the IPSs versus norm, with the result displayed in Fig. \ref{iso}(b). It demonstrates that the size of the IPSs decreases as the increasing of total norm. Typical example of an IPS is displayed in Fig. \ref{iso}(c). On the other hand, in Fig. \ref{Char}(c2) , we find that the value of $\mu$ approaches $0$ when $C_{\mathrm{V}}$ is larger than the value that can produce IPSs[see Fig. \ref{Char}(b2)].

\section{Motion, collision and rotation of the solitons}
\subsection{Motion and collision of the solitons}
\begin{figure}[tbp]
\centering%
{\label{fig5a}
\includegraphics[scale=0.45]{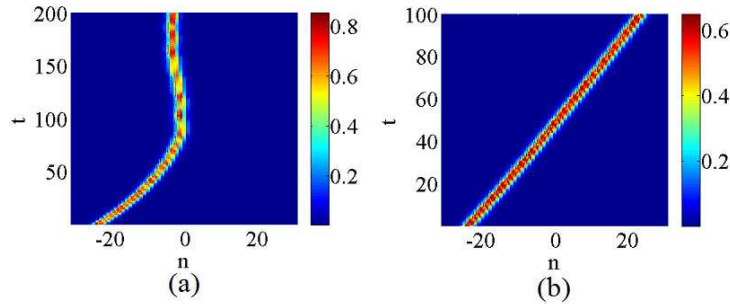}}%
\caption{(Color online) (a) Example of the immobile soliton with $(C_{\mathrm{H}},C_{\mathrm{V}}, P)=(1,2.9,6)$. Here the strength of kick is set as $\eta=0.17$. (b)Example of the mobile soliton with $(C_{\mathrm{H}},C_{\mathrm{V}}, P)=(1,3.08,6)$. Here the strength of kick is set as $\eta=0.17$. Note that these two panels are the cross-section of $n (\mathrm{vertical})-t$ plane at $m=0$. In these two panels, $\sigma=0$, and the dipoles in the system are orientated to the $m$ (horizontal) direction. }
\label{motionExp}
\end{figure}
\begin{figure}[tbp]
\centering%
{\label{fig6a}
\includegraphics[scale=0.3]{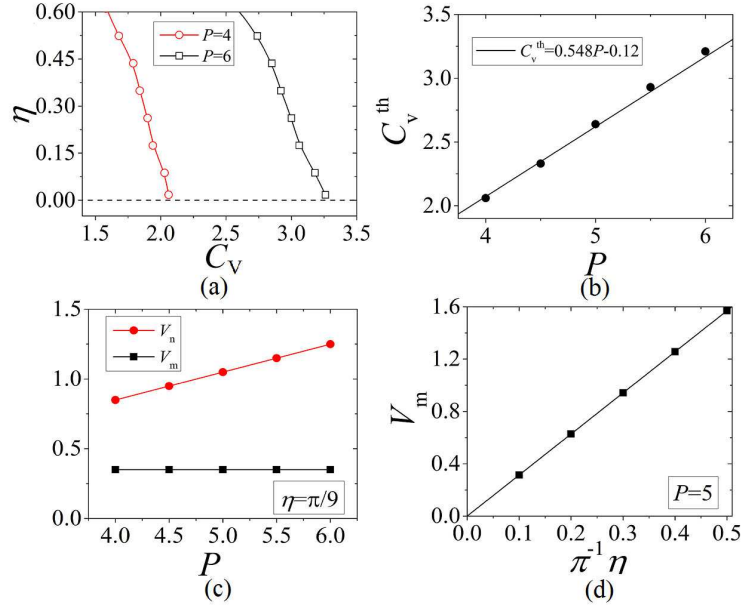}}%
\caption{(Color online) (a)Critical curves for the mobile (left side of the curves) and immobile (right side of the curves) solitons with $P=4$, and $P=6$, respectively. (b)The threshold value of the vertical hopping rate, $C^{\mathrm{th}}_{\mathrm{V}}$ for the IPS, as a function of $P$. The specific relation between $C^{\mathrm{th}}_{\mathrm{V}}$ and $P$ is perfectly fitted by a linear function $C^{\mathrm{th}}_{\mathrm{V}}=0.548P-0.12$. (c) Moving velocities $V_{m,n}$ of IPSs vs $P$, which are measured by the displacement divided by $t$. (d) The horizontal velocity $V_{m}$ of an IPS vs $\eta$ with $P=5$. In all these cases, $C_{\mathrm{H}}\equiv1$, $\sigma=0$, and all dipoles are directed to the $m$ (horizontal) direction.  }
\label{motionChar}
\end{figure}

Mobility of discrete solitons across the discrete lattices is a basic issue in the theory of the discrete nonlinear dynamic \cite{kevrekidis}. When the solitons are narrow and strongly pinned to the underlying lattice, the discrete solitons become immobile \cite{Kuzetsov}. While the size of solitons is sufficiently large, the discrete system can be treated as a quasicontinued system, and therefore the solitons can become mobile \cite{Jinhong,Zhenjing}. This can be partially explained by the Ablowitz-Ladik model \cite{Kivshar2,Konotop,Cai}. Generally, the motion of discrete localized modes across the lattice can be interpreted in terms of the PN barrier \cite{Kivshar,Morandotti}, which is an energy difference between the lattices. It is possible to obtain a mobile solitary state as long as the PN barrier is overcome. In a 1D setting, it has already been demonstrated that the attractive and repulsive DDIs could help to reduce and enhance the PN barrier \cite{Gligoric,Gligoric2,Zhihuan}, respectively. In our 2D discrete setting, the polarization of the dipoles might give rise to soliton motion that exhibits anisotropic behaviors because of the anisotropic DDIs \cite{Yongyao3}. Thus, in this subsection, we still assume that the dipole polarization is along the $m$ (horizontal) direction. As discussed above, the DDIs in the horizontal and vertical directions are attractive and repulsive, respectively. Under this circumstance, in the horizontal direction, the attractive DDIs can help to reduce the PN barrier, making it possible to kick the solitons moving freely in this direction; while in the vertical direction, due to the enhancement of PN barrier by the repulsive DDIs, it is nontrivial to study the motion of the solitons in this direction by tuning $C_{\mathrm{V}}$. Therefore, in the following, we solely study the motion of the solitons in the vertical direction. In this case, the initialized solitons by applying a kick (in the vertical direction) is:
\begin{eqnarray}
\psi_{mn}(t=0)=\phi_{mn}e^{i\eta n}, \label{kickinitial}
\end{eqnarray}
where $\eta$ is the strength of the kick and $\phi_{mn}$ is a stationary solution (recall that the vertical direction is the $n$-direction). Numerical simulations demonstrate that, for the fixed value of $P$ and $C_{\mathrm{V}}$ (note that $C_{\mathrm{H}}$ has been fixed to 1), the solitons can be set into motion when the kick exceeds a finite threshold value $\eta_{c}$. If the kick $\eta<\eta_{c}$ , the solitons may only be shifted from the initial position by a finite distance, i.e., the solitons start to move but then come to a halt. Figures \ref{motionExp}(a) and \ref{motionExp}(b) show two typical examples of the dynamic of the kicked solitons when the strength of kick is smaller and larger than $\eta_{c}$, respectively. Figure \ref{motionChar}(a) shows $\eta_{c}$ as a function of $C_{\mathrm{V}}$ for two different values of $P$. In this panel, it is seen that there is a threshold value of the hopping rate, $C^{\mathrm{th}}_{\mathrm{V}}$, satisfying $\eta_{c}\equiv0$. When $C_{\mathrm{V}}>C^{\mathrm{th}}_{\mathrm{V}}$, the solitons can be kicked to move with an arbitrarily small $\eta$. In this case, the size of solitons is sufficiently wide, hence making the discrete system retreat to a quasicontinued system. As a result, the solitons can move freely after imposing an arbitrary kick. Figure \ref{motionChar}(b) displays $C^{\mathrm{th}}_{\mathrm{V}}$ as a function of $P$, showing that the value of $C^{\mathrm{th}}_{\mathrm{V}}$ increases linearly as $P$.

The comparison between Fig. \ref{iso}(b) and Fig. \ref{motionChar}(b) shows that the IPSs can be found with the condition of $C_{\mathrm{V}}>C^{\mathrm{th}}_{\mathrm{V}}$, indicating that the IPSs can move freely in the 2D space. To study the motion of the IPSs, they are kicked in both (the horizontal and the vertical) directions with the same $\eta$ and their propagating velocity ( the displacement divided by $t$) is measured. Figure \ref{motionChar}(c) shows the measured velocity in both directions versus $P$ for the fixed $\eta$. In this panel, we can see that the horizontal velocity, $V_{m}$, does not change with $P$, and the vertical velocity, $V_{n}$, satisfies
\begin{eqnarray}
V_{n}=C_{\mathrm{V}}(P)V_{m}. \label{Vn}
\end{eqnarray}
Equation (14) indicates that the motion of the IPSs is anisotropic.
The relation between $V_{m}$ and $\eta$ is also investigated, with the results shown in Fig. \ref{motionChar}(d), depicting clearly the linear relationship of $V_{m}\equiv\eta$. This relation can be naturally understood by the conservation of momentum. However, according to Eq.(\ref{Vn}) and the conservation law of momentum, it is found that the effective mass of the IPSs is anisotropic with respect to different value of $C_{\mathrm{V}}$.

\begin{figure}[tbp]
\centering%
{\label{fig7a}
\includegraphics[scale=0.4]{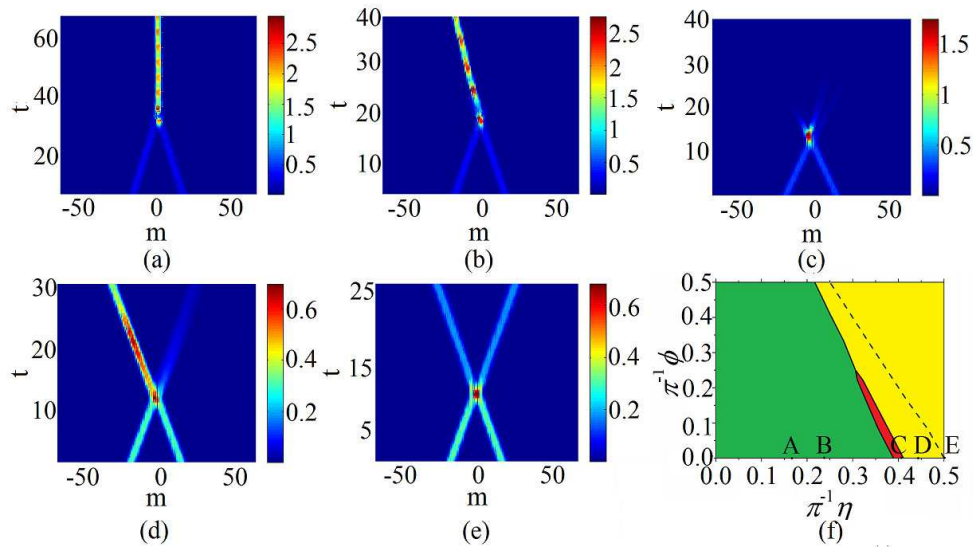}}%
\caption{(Color online) Typical examples of collisions between IPSs with $P=5$, $C_{\mathrm{H}}=1$, and $C_{\mathrm{V}}=3$. Panels (a-e) are shown in the cross section of $n=0$, corresponding to the cases of points A-E shown in panel (f). The distance between two initially composed solitons is $2m_{0}=32$. (a) Symmetric merger with $(\eta,\phi)=(\pi/6,0)$. (b) Asymmetric merger with $(\eta,\phi)=(11\pi/36,0)$. (c) Full destruction with $(\eta,\phi)=(73\pi/180, 0)$. (d, e) Half-destruction with $(\eta,\phi)=(4\pi/9,0)$ and completely elastic collision with $(\eta,\phi)=(\pi/2,0)$, respectively.  (f) Plane of $(\eta,\phi)$ (with $\eta,\phi\in[0,\pi/2]$) showing the outcomes of collisions between IPSs. The green area, red area, yellow area, and the dashed line of panel (f) denote four types of outcomes of collisons: the merger, full destruction, half destruction and completely elastic collision, respectively. In all these cases, $C_{\mathrm{H}}\equiv1$, $\sigma=0$, and all dipoles are directed to the $m$ (horizontal) direction.}
\label{collision}
\end{figure}


The robust mobility of the IPSs in the 2D anisotropic lattices offers an opportunity to study the collision between them \cite{PaPa,Dmitriev,Khaykovich}. Here, their collision in the horizontal ($m$) direction is considered. To this end, we numerically simulated Eq. (\ref{GP}) by inputting two composed solitons that are separated by a distance $2m_{0}$ [their centers are placed at sites ($\mp m_{0},0$)] with a phase shift $\phi$ between them. Actually, in addition to the parameter $\eta$ that determined the relative velocity of IPS, this phase difference $\phi$ is another controllable parameter for the collision of solitons. In this case, the initial state of the solitons is written as
\begin{eqnarray}
\psi_{mn}(t=0)=\phi_{-m_{0},0}e^{i\eta m}+\phi_{m_{0},0}e^{-i(\eta m-\phi)}.
\end{eqnarray}
In simulations, $m_{0}$ is selected as 16.  The norm of each IPS is selected as $P=5$, which requires $(C_{\mathrm{H}}, C_{\mathrm{V}})=(1,3)$. Based on these settings, the moving IPSs feature four types of collision showing in the $(\eta,\phi)$ plane; see Fig. \ref{collision}(f): merger (green color area which includes symmetric and asymmetric merger), half-destruction (yellow color area), full-destruction ( red color area), and completely elastic collision (dashed line). Particular examples of these types of collision are displayed in Figs. \ref{collision}(a)-\ref{collision}(e), respectively, corresponding to the points A-E shown in Fig. \ref{collision}(f), respectively. It is necessary to point out that even though the result shown in Fig. \ref{collision}(b) is similar to that shown in \ref{collision}(d), they are essentially different. In Fig. \ref{collision}(b), two solitons merge together after collision, therefore, we can see that the intensity of the merger is approximately twice to the initial intensity of a soliton before the collision. However, in Fig. \ref{collision}(d), the total density after soliton collision is divided into two parts: one survives and the other decays with radiation. The intensity of the survived one is approximately equal to the intensity of one of the initial soliton before the collision.

With the consideration of DDIs, our numerical simulations demonstrate that, in most cases, the collisions between two IPSs are inelastic. The completely elastic collision only occurs when $\phi=-2(\eta-\pi/2)$ for $\eta\in[0,\pi/2]$. The soliton collision with different initial distances and different value of $P$ are also simulated, but no other type of outcome can be found.

\subsection{Rotation of the lattice solitons}
\begin{figure}[tbp]
\centering%
{\label{fig8a}
\includegraphics[scale=0.2]{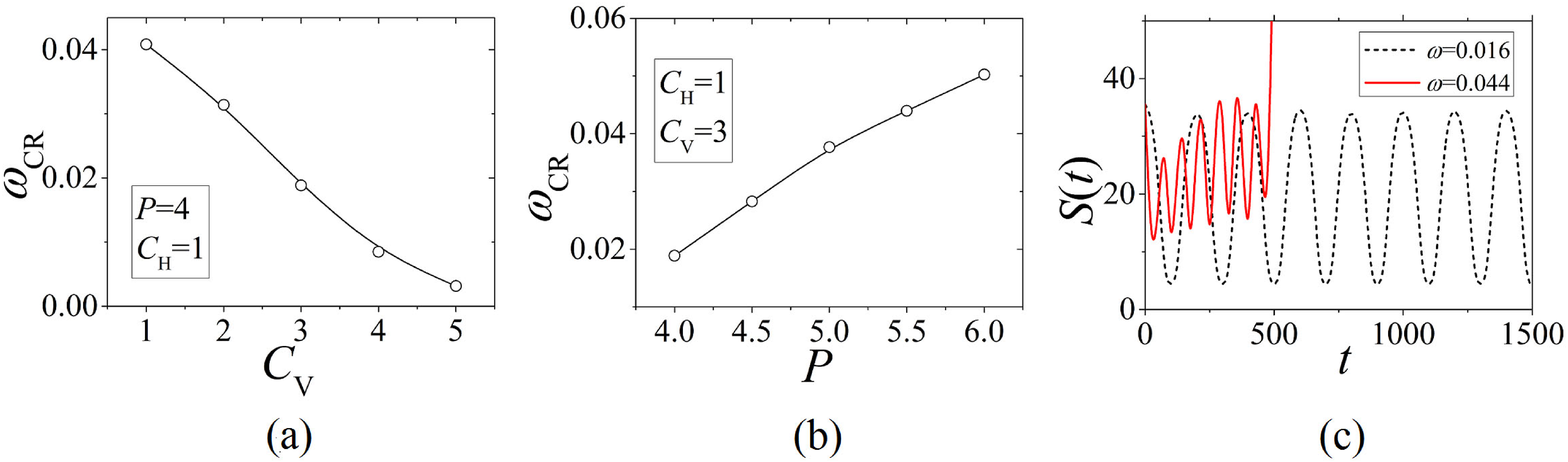}}%
\caption{(Color online) (a,b) $\omega_{\mathrm{CR}}$ as functions of $C_{\mathrm{V}}$ and $P$, respectively. (c) Dynamic of the effective area of the localized state, which is defined as $S(t)=\left(\sum_{mn}|\psi_{mn}(t)|^{2}\right)^{2}/\sum_{mn}|\psi_{mn}(t)|^{4}$ for $\omega=0.016$ (black dash curve) and $\omega=0.044$ (red solid curve). These results are obtained with the initial conditions: the norm of IPS is $P=5$, $\left(C_{\mathrm{H}}, C_{\mathrm{V}}\right)=(1,3)$, and $\sigma=0$.}
\label{omega}
\end{figure}

\begin{figure}[tbp]
\centering%
{\label{fig9a}
\includegraphics[scale=0.35]{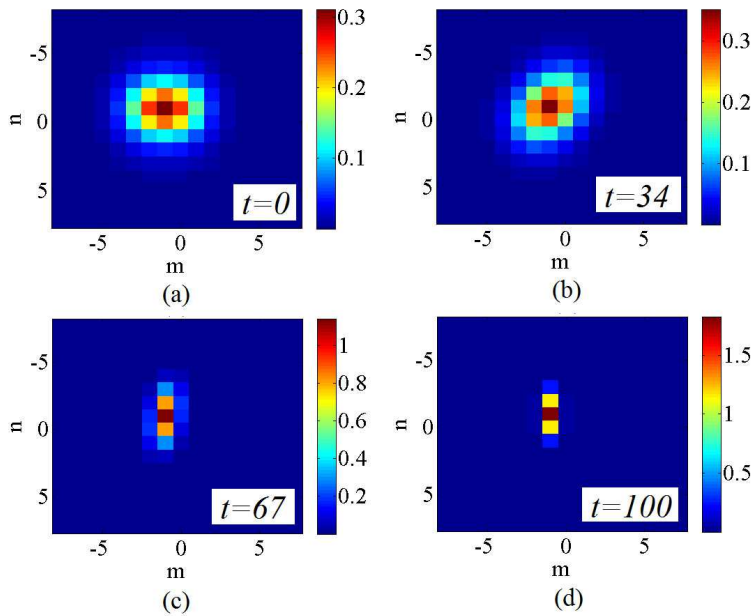}}%
\caption{(Color online) Stable rotation of a lattice IPS driven by an external rotating magnetic field, with the same case as shown in the black dash curve in Fig. \ref{omega}(c). (a) $t=0$, i.e. $\theta(t)=0^{\circ}$. (b) $t=34$, i.e. $\theta(t)\approx30^{\circ}$. (c) $t=67$, i.e. $\theta(t)\approx60^{\circ}$. (d) $t=100$, i.e. $\theta(t)=90^{\circ}$. Notice that in all these cases, $\sigma$ is fixed to 0.}
\label{omegaExp}
\end{figure}

Compared to the BECs achieved by the non-polar atoms, the advantages of dipolar BECs are that they are easier to be rotated by an external rotational field \cite{Tikhonenkov,Xunda}, in which case the magnetic moments are polarized and rotated following the external magnetic field. The rotational dynamics of the dipolar matter-wave solitons have already been studied in 2D free space \cite{Tikhonenkov,Xunda}. In that case, solitons can remain localized during the ratation. However, in the OLs, especially in the anisotropic OLs, due to the strongly combined effect between the anisotropic linearity and nonlocal nonlinearity, the size of the solitons is expected to be affected significantly by the rotation of dipoles, making it possible to become a breather which is difficult to keep localized. Therefore, it is nontrivial to find the existence of a threshold, below which the breather can survive.

Such dynamical process can be described by Eq. (\ref{GP}) by replacing $\theta$ to
\begin{eqnarray}
\theta'=\theta+\omega t, \label{theta2}
\end{eqnarray}
where $\theta$ is defined by Eq. (\ref{theta}). Numerical simulations demonstrate that the threshold of $\omega$ (namely $\omega_{\mathrm{CR}}$), below which the breathers can remain localized, is a function of $C_{\mathrm{V}}$ and $P$; see Figs. \ref{omega}(a and \ref{omega}(b), respectively. Numerical results in Figs. \ref{omega}(a and \ref{omega}(b) suggest that the breather can only survive for small value of $\omega$. The breathers, containing larger norms and being in the lattices with smaller value of $C_{\mathrm{V}}$, are shown to be able to survive easily in the stir of the external rotating field.  Typical example of this localized dynamic is depicted by the black dash curve of $S(t)$ in Fig. \ref{omega}(c) with the initial condition of $(P, C_{\mathrm{H}}, C_{\mathrm{V}})=(5,1,3)$ and $\omega=0.016$. The patterns of the breathers at $\theta=0^{\circ}$, $30^{\circ}$, $60^{\circ}$ and  $90^{\circ}$  are displayed in Fig. \ref{omegaExp}. While $\omega>\omega_{\mathrm{CR}}$, breathers decay with spreading out. Typical example of this dynamic is also presented shown by the red solid curve of $S(t)$ in Fig. \ref{omega}(c), which starts from the same initial conditions, but with $\omega=0.044$. The delocalization of the soliton occurs after $t\approx500$.

\section{Influence of the contact nonlinearity}
\begin{figure}[tbp]
\centering%
{\label{fig10a}
\includegraphics[scale=0.35]{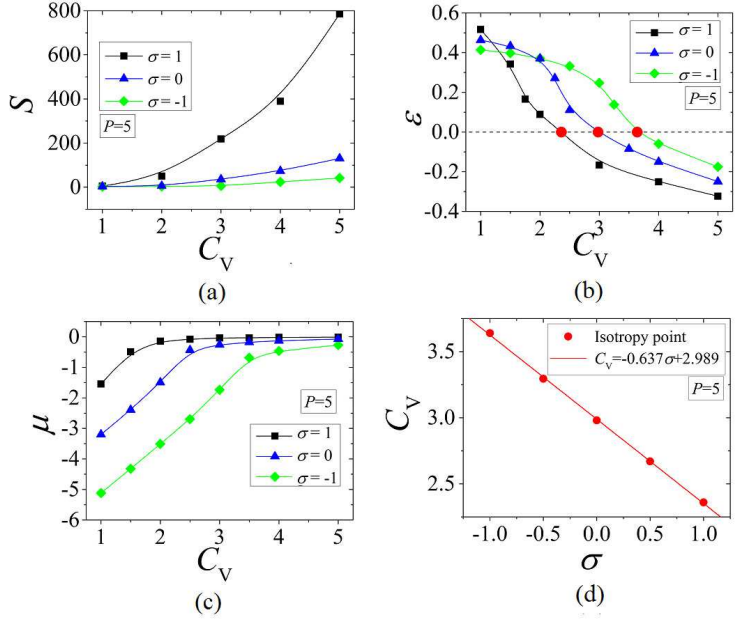}}%
\caption{(Color online) (a,b,c) The effective area $S$ (a), the degree of anisotropic coefficient $\varepsilon$ of the solitons (b), and the chemical potential $\mu$ (c) versus $C_{\mathrm{V}}$ for $\sigma=1$ (square line), 0 (triangle line) and $-1$ (rhombus line), respectively. Here, the total norm of the solitons is fixed to $P=5$ and the horizontal hopping rate $C_{\mathrm{H}}$ is fixed to 1. (d) The relationship between $C_{\mathrm{V}}$ and $\sigma$ for the formations of IPSs can be linearly fitted by $C_{\mathrm{V}}=-0.637\sigma+2.989$. Here we fixed $C_{\mathrm{H}}\equiv1$ and $P=5$. In all these cases, the dipoles assumed to be orientated to the $m$ (horizontal) direction. }
\label{gchar}
\end{figure}

\begin{figure}[tbp]
\centering%
{\label{fig11a}
\includegraphics[scale=0.28]{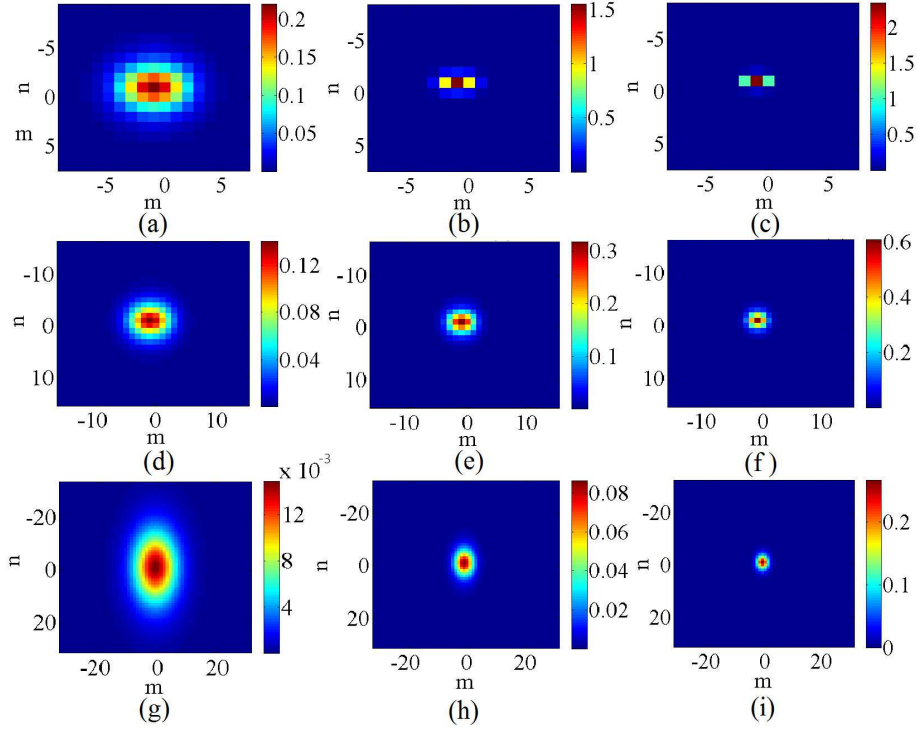}}%
\caption{(Color online) (a,b,c) Horizontally-elongated solitons with $\sigma=1$, 0 and -1, respectively. (d,e,f) IPSs with $\sigma=1$, 0 and -1, respectively. (g,h,i) Vertically-elongated solitons with $\sigma=1$, 0 and -1, respectively. Here, the total norm of these solitons is select as $P=5$. The numerical domain we used is set as $128\times128$. Here the regions of the above panels are drawn according to the size of the solitons. In all these cases, the dipoles assumed to be orientated to the $m$ (horizontal) direction.  }
\label{gExp}
\end{figure}

The influence of the contact and/or local nonlinearity to the dipolar matter-wave lattice solitons in such a linearly anisotropic lattices is considered in this section. In Eq. (\ref{GP}), the contact nonlinearity is contributed from the parameter $\sigma$. $\sigma<0$ means that the contact nonlinearity is self-attractive; while $\sigma>0$ means that the contact nonlinearity is self-repulsive. Here, we only consider the case of fundamental unstaggered bright solitons, with the total norm of the solitons fixed to 5. Figures \ref{gchar}(a)-\ref{gchar}(c) display the typical examples of the effective area $S$, anisotropic degree $\varepsilon$, and the chemical potential $\mu$ versus $C_{\mathrm{V}}$ with different values of $\sigma$. It can be seen from Figs. \ref{gchar}(a)-\ref{gchar}(c)  that the solitons with $\sigma=1$ are the most sensitive to the variation of $C_{\mathrm{V}}$. This is because the self-repulsion helps the spreading of the fundamental solitons. Therefore, with the increasing of $C_{\mathrm{V}}$, their elongated property of the solitons with $\sigma=1$ is changed more significantly than that of other cases. Further, the relationship between $C_{\mathrm{V}}$ and $\sigma $ for the formations of IPSs is displayed in Fig. \ref{gchar}(d). In this panel, this relationship is perfectly fitted by a linear function. Within the regime below the line, the solitons are horizontally-elongated, while above the line, the solitons are vertically-elongated. Typical examples of the solitary patterns with $\sigma=1$, 0 and $-1$ for horizontally elongated solitons, IPSs and vertically elongated solitons are displayed in Fig. \ref{gExp}.

\section{Conclusion}

The objective of this work was to study 2D solitons in the disk-shaped dipolar BECs trapped in a strongly anisotropic OL potential in the disk's plane. The OL is a square lattice, which can be formed by interfering two pairs of plane waves with different intensity, hence the hopping rates of the condensates between the adjacent lattices in the horizontal and vertical directions are different. This would lead to a linearly anisotropic OL. We assumed that the dipole polarization is parallel to the 2D plane and directed to the $x$ (horizontal) direction by an external static magnetic field, therefore the long-range DDIs are anisotropic. This gives rise to the nonlocal anisotropic nonlinearity. The setting considered here can be described by the 2D discrete GPE with different hopping rates along the $m$ (horizontal) and the $n$ (vertical) directions, respectively, due to the fragmentation of the condensates by the strong anisotropic OL.

The pattern, motion, collision and rotation of 2D fundamental unstaggered lattice solitons were studied by means of numerical simulations under the influence of the linearly and nonlinearly anisotropy of the system. It was found that the linear anisotropy of the system can enhance, reduce and even change the profiles of the solitons. Typically, IPSs were found when these two types of anisotropic effects reach a balance. The existence of such IPSs was identified by the hopping rate which is perpendicular to the polarized orientation of the dipoles (i.e., $C_{\mathrm{V}}$), and the total norm (i.e., $P$). For the dynamics, we investigated the influence of such combined effects (linearly and nonlinearly anisotropic effects) to the motion, collision and rotation of the IPSs. It was found that the IPSs can move freely in both directions, and the collision between two IPSs was also studied in detail. For the rotation of the IPSs, a critical angular velocity was found, below which the IPSs play as breathers following the rotation of the external magnetic field that determines the polarization of the atomic magnetic moments. Finally the influences of the contact nonlinearity to the 2D lattice solitons were also considered, and the relationship between the vertical hopping rate and the strength of contact nonlinearity at which the IPSs are found, was explored too.

The present analysis can be extended in some directions. Firstly, Refs. \cite{Gligoric4} and \cite{Kevredkidis2005,Mayteevarunyoo,CGH} suggest that single anisotropic DDIs and single anisotropic lattice can support families of stable vortex solitons, respectively. It is challenged to consider the stabilities of such families of vertex solitons after considering the combined effects of the linearly and nonlinearly anisotropy. Further, it may be interesting to consider the IPSs in another type of anisotropic nonlinearity formed by electric quadrupolar BECs with long-range quadrupole-quadrupole interactions (QQIs) \cite{Yongyao3,Jiasheng}, which are more anisotropic and tighter than the DDIs. Finally, a challenging option is to seek for stable solitons in the 3D anisotropic setting.

\begin{acknowledgments}
Y.L. appreciate useful discussions with Prof. Boris A. Malomed (Tel-Aviv University) and Shenhe Fu (Sun Yet-Sen University). This work was supported by the National Natural Science Foundation of China through Grants No. 11575063.
and 11204089
\end{acknowledgments}

\bibliographystyle{plain}
\bibliography{apssamp}

\end{document}